\documentclass[aps,prl,twocolumn,groupedaddress]{revtex4-1}
\usepackage{graphicx}
\usepackage{dcolumn}
\usepackage{bm}
\usepackage{amsmath}
\usepackage{amssymb}
\usepackage{latexsym}
\usepackage{epsfig}
\usepackage{amsbsy}
\usepackage{array}
\usepackage{amssymb}
\usepackage{setspace}
\usepackage{bm}
\usepackage{xcolor}

\def\sint{\ifmmode{- \!\!\!\!\!\! \int}
    \else{\hbox{$- \!\!\!\! \int \ $}}\fi}

\begin{document}

\title{Topological Charge/spin density Wave in InAs/GaSb Quantum Wells under an In-plane Magnetic Field}

\author{Lun-Hui Hu$^{1,2,3}$}
\author{Chih-Chieh Chen$^{1,3}$}
\author{Chao-Xing Liu$^{2}$}
\email{cxl56@psu.edu}
\author{Fu-Chun Zhang$^{4,3}$}
\author{Yi Zhou$^{1,3}$}

\affiliation{$^1$Department of Physics, Zhejiang University, Hangzhou, Zhejiang, 310027, China}
\affiliation{$^2$Department of Physics, The Pennsylvania State University, University Park, Pennsylvania 16802}
\affiliation{$^3$Collaborative Innovation Center of Advanced Microstructures, Nanjing 210093, China}
\affiliation{$^4$Kavli Institute for Theoretical Sciences,  University of Chinese Academy of Sciences, Beijing 100190, China}

\date{\today}

\begin{abstract}
  We study interaction effect of quantum spin Hall state in InAs/GaSb quantum wells under an in-plane magnetic field by using the self-consistent mean field theory. We construct a phase diagram as a function of intra-layer and inter-layer interactions, and identify two novel phases, a charge/spin density wave phase and an exciton condensate phase. The charge/spin density wave phase is topologically non-trivial with helical edge transport at the boundary, while the exciton condensate phase is topologically trivial. The Zeeman effect is strongly renormalized due to interaction in certain parameter regimes of the system, leading to a much smaller $g$-factor, which may stabilize the helical edge transport.
\end{abstract}


\maketitle

{\it Introduction -}
Interaction effect plays an intriguing role in topological physics \cite{hasan_rmp_2010,qi_rmp_2011}, which has not been well explored, particularly in realistic topological materials. Theoretically, it was predicted that interaction can stabilize or destroy topological states, and even enable new topological classifications \cite{fidkowski_kitaev_prb_2010,fidkowski_kitaev_prb_2011,ryu_prb_2012,qi_njp_2013,yao_prb_2013,fidkowski_prx_2013, gu_prb_2014,wang_prb_2014,you_arxiv_2014,you_prb_2014}. Experimentally, interaction effect is much less understood since only a few topological systems, including InAs/GaSb quantum wells \cite{li_prl_2015} and topological Kondo insulator SmB$_6$ \cite{neupane_natcomm_2013,Jiang_natcomm_2013,xu_prb_2013,Xia_scirep_2014,xu_arxiv_2014, neupane_prl_2015}, are known to possess strong interactions. Recent experiments in InAs/GaSb type-II quantum wells, a two dimensional quantum spin Hall insulator \cite{liu_prl_2008,knez_prl_2011,knez_prl_2014,fabrizio_prl_2014,du_prl_2015,fanming_prl_2015,susanne_prb_2015, du_li_arxiv_2016}, have shown that the temperature dependence of helical edge transport follows the transport behavior of Luttinger liquids with Luttinger parameter $K\sim 0.21$ \cite{li_prl_2015}, indicating strong repulsive Coulomb interaction
in this topological system \cite{wu_prl_2006,joseph_prl_2009}. Previous theoretical studies \cite{pikulin_prl_2014} also proposed the possibility of
a novel topological exciton condensation (EC) phase in this system. Thus, InAs/GaSb quantum wells provide us a platform to explore interaction effect in realistic topological materials.

A puzzling observation in previous experimental studies of InAs/GaSb quantum wells is that helical edge transport is extremely robust under an in-plane magnetic field and the quantized conductance plateau persists up to 12 Tesla magnetic field \cite{du_prl_2015,du_arxiv_2015}. Theoretically, magnetic fields are expected to break time reversal symmetry, thus leading to backscattering in helical edge transport \cite{joseph_prb_2010,pikulin_prb_2014}. Given that the importance of Coulomb interaction in this system, this motivates us to study the interaction effect in InAs/GaSb quantum wells under an in-plane magnetic field. In this work, we extract the phase diagram as a function of inter-layer $V$ and intra-layer $U$ interaction strengths based on the mean field theory, and identify two distinct interacting phases, a EC phase and a charge/spin density wave (CDW/SDW) phase. In particular, we find that EC phase is topologically trivial while helical edge states are supported at the boundary for the CDW/SDW phase when spin conservation is presence. Strong correction to the Zeeman effect (either reduction or enhancement) is found for the topological CDW/SDW phases, depending on detailed material parameters. Thus, our results provide a possible scenario to understand the robust helical transport under in-plane magnetic fields.

\begin{figure}[!htbp]
   \centering
   \includegraphics[width=3.2in]{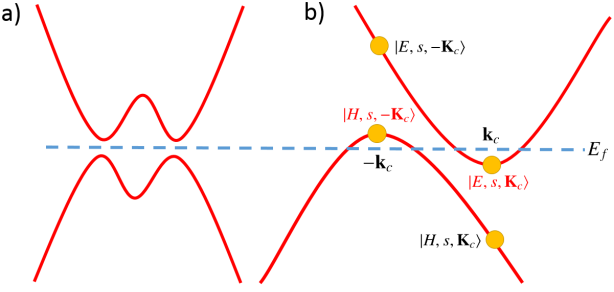}
   \caption{\label{fig-1-sketch} Schematics of the inverted band structure in the InAs/GaSb quantum wells (a) at a zero magnetic field and (b) at a finite in-plane magnetic field. The pairing between $\vert E,s,\mathbf{k}_c\rangle$ and $\vert H,s,-\mathbf{k}_c\rangle$ is indirect exciton condensate (EC) order parameter. The pairing between $\vert E/H, s,\mathbf{k}_c\rangle$ and $\vert E/H, s,-\mathbf{k}_c\rangle$ is density wave order parameter.}
\end{figure}

{\it Model Hamiltonian - }
In type-II InAs/GaSb quantum wells, electrons are confined in the InAs layers while holes are localized in the GaSb layers, thus forming a bilayer layer electron-hole system, which can be described by the Bernevig-Hughes-Zhang (BHZ) model \cite{bernevig_prl_2006,Bernevig_science_2006,liu_prl_2008,rothe_review_njp_2010} with four bands (two spin-split electron bands and two spin-split hole bands). We consider an in-plane magnetic field along the y direction $\mathbf{B}_{\parallel}=B_{\parallel}\vec{e}_y$. We first focus on the orbital effect and the influence of Zeeman effect will be discussed later. The orbital effect of the in-plane magnetic field is normally not important for a two dimensional system. However, in our case, since electron and hole bands are separated into two layers by around 10 nm of spacing, the orbital effect of in-plane magnetic fields can induce an opposite shift between the electron and hole bands \cite{yang_prl_1997}. To describe this effect, we choose the Landau gauge $\mathbf{A}=B_{\parallel}z\vec{e}_x$ and set a middle point between the two layers as the origin of the coordinate system \cite{hu_prb_2016}. Therefore, the BHZ Hamiltonian with the orbital effect of in-plane magnetic fields reads,
\begin{align}
  \mathcal{H}_{\text{BHZ}}(\mathbf{k}) &= \left(M-B(\mathbf{k}-\mathbf{k}_c)^2\right) s_0 \otimes(\sigma_0+\sigma_z)/2  \nonumber \\
       & + \left(-M+B(\mathbf{k}+\mathbf{k}_c)^2\right) s_0 \otimes(\sigma_0-\sigma_z)/2 \nonumber \\
       & + (Ak_x)s_z\otimes\sigma_x + (Ak_y)s_0\otimes\sigma_y
\end{align}
under the basis $\{E\uparrow, E\downarrow,H\uparrow,H\downarrow\}$, where the Pauli matrix $s$ is for pseudospin $\{\uparrow,\downarrow\}$ and $\sigma$ gives band $\{E,H\}$, and the momentum shift $\mathbf{k}_c=(\phi_0,0)$ with $\phi_0=\frac{e}{\hbar}\frac{d}{2}B_{\parallel} $, where $ d $ is the distance between electron gas in InAs layer and hole gas in GaSb layer. At zero magnetic field $B_{\parallel}=0$, the energy dispersion of InAs/GaSb quantum wells is inverted with a small hybridization gap, as shown in Fig. \ref{fig-1-sketch}(a). With a finite magnetic field, the system is driven into a semi-metal phase \cite{yang_prl_1997,hu_prb_2016} due to the opposite shift between the electron and hole bands, as shown in Fig. \ref{fig-1-sketch}(b). The discussion below is focused on the interaction effect on this semi-metal phase. The parameters are chosen as $B_{\parallel}=6$ Tesla and $d=10$ nm, yielding $\phi_0=0.0456$ nm$^{-1}$. Other parameters in the BHZ model, such as $B$, $M$ and $A$, are taken from the Ref. \cite{franz_book_2013}.

Next we consider the electron-electron Coulomb interaction \cite{kotov_rmp_2012}, given by
\begin{align}\label{eq-coulomb-total}
   \mathcal{H}_{I} = \sum_{a,a'} \sum_{\mathbf{k},\mathbf{k}',\mathbf{q}} V^{aa'}(q) \hat{C}_{a,\mathbf{k}}^\dagger \hat{C}_{a',\mathbf{k}'}^\dagger C_{a',\mathbf{k}'+\mathbf{q}} C_{a,\mathbf{k}-\mathbf{q}} ,
\end{align}
where $a\in\{E\uparrow,E\downarrow,H\uparrow,H\downarrow\}$. Since electron and hole bands are separated at two layers, the interaction $V^{aa'}(q)$ can be characterized by two terms, the inter-layer interaction $V^{E_s, H_{s'}}=V e^{-qd}/q$ and the intra-layer interaction $V^{E_s, E_{s'}}=V^{H_s, H_{s'}}=U/q$ with $s,s'$ for spin. Here $V$ and $U$ are in unit of $V_0=\frac{e^2}{2\epsilon_0 (2\pi)^2}$ (meV$\cdot$nm).

We have shown in Fig.~\ref{fig-1-sketch}(b) that under an in-plane magnetic field, the energy dispersion
for the BHZ model can possess one electron and one hole Fermi pocket. The Coulomb interaction can induce scattering between electron and hole Fermi pockets to open a gap. Thus, we study possible insulating phases induced by Coulomb interaction based on the mean field approximation \cite{naveh_prl_1996}, as discussed in details in the appendix.
The mean field decomposition of Eq.~\eqref{eq-coulomb-total} is taken as
\begin{align}
     \mathcal{H}_{I} &= \sum_{a,a'} \sum_{\mathbf{k},\mathbf{k}'} V^{aa'}(\vert \mathbf{k}-\mathbf{k}' \vert) \hat{C}_{a,\mathbf{k}}^\dagger \hat{C}_{a',\mathbf{k}'\pm 2\mathbf{k}_c}^\dagger C_{a',\mathbf{k} \pm 2\mathbf{k}_c} C_{a,\mathbf{k}'}  \nonumber \\
   & \rightarrow \sum_{a,a'} \sum_{\mathbf{k}} \bar{\Delta}^{a,a'}(\mathbf{k})  \hat{C}_{a,\mathbf{k}}^\dagger C_{a',\mathbf{k} \pm 2\mathbf{k}_c} ,
\end{align}
where we have defined the order parameter as $\Delta^{a,a'}(\mathbf{k}') = \left\langle \hat{C}_{a',\mathbf{k}' \pm 2\mathbf{k}_c}^\dagger  C_{a,\mathbf{k}'} \right\rangle$ and the corresponding 8-by-8 gap function as $ \bar{\Delta}^{a,a'}(\mathbf{k}) = -\sum_{\mathbf{k}'} V^{aa'}(\vert \mathbf{k}-\mathbf{k}' \vert)  \Delta^{a,a'}(\mathbf{k}') $.
Based on the above decomposition, the mean field Hamiltonian, which is expanded
around $k_c$ for electron Fermi pocket and around $-k_c$ for hole Fermi pocket, is written as
\begin{align}\label{eq-ham-mf}
  \mathcal{H}_{\text{MF}} = \sum_{\bar{\mathbf{k}}} \left\lbrack\begin{array}{cc}
                           \mathcal{H}_{\text{BHZ}}(\bar{\mathbf{k}}+\mathbf{k}_c) & 0 \\
                           0 &   \mathcal{H}_{\text{BHZ}}(\bar{\mathbf{k}}-\mathbf{k}_c)
                         \end{array}\right\rbrack  + \mathcal{H}_{\bar{\Delta}}(\bar{\mathbf{k}}),
\end{align}
where $\bar{\mathbf{k}}$ is a small momentum $\bar{\mathbf{k}}\ll k_c$. The 8-by-8 gap function can be expanded within the 16 independent s-wave real order parameters \cite{budich_prl_2014}, labelled as $\mathcal{D}_{\alpha\beta x}=s_\alpha\otimes\sigma_\beta\otimes\tau_x$ or $\mathcal{D}_{\alpha\beta y}=s_\alpha\otimes\sigma_\beta\otimes\tau_y$, where $\tau$ is the Pauli matrix for valley degree. Thus, the mean-field Hamiltonian reads $\mathcal{H}_{\bar{\Delta}}(\bar{\mathbf{k}})= \sum_{\alpha,\beta}\bar{\Delta}_{\alpha\beta \tau}(\bar{\mathbf{k}}) \mathcal{D}_{\alpha\beta\tau}$ with $\tau=x$ or $y$. In this work, higher order term of order parameters, such as $\left\langle \hat{C}_{a',\mathbf{k}' \pm 2m\mathbf{k}_c}^\dagger  C_{a,\mathbf{k}'} \right\rangle$ with $m=\{2,3,4,\cdots\}$, is neglected. Similar to bilayer HgTe system \cite{budich_prl_2014}, we will only focus on four different order parameters among these 16 real order parameters, namely
$\mathcal{D}_{xxx}$ and $\mathcal{D}_{yxy}$, which can be induced by the inter-layer interaction $V$, and
$\mathcal{D}_{0zx}$ and $\mathcal{D}_{zzx}$, which result from the intra-layer interaction $U$. All of them can gap out both the electron and hole Fermi pockets, and thus are energetically favorable. We also notice that $\mathcal{D}_{xxx}$ and $\mathcal{D}_{yxy}$ can be explicitly written in the form of
$\left\langle \hat{C}_{\mathbf{k}, \sigma ,s}^\dagger \hat{C}_{\mathbf{k}+2\mathbf{k}_c,\sigma',s'} \right\rangle$, thus physically representing EC order parameters in differen spin channels. On the other hand, $\mathcal{D}_{0zx}$ and $\mathcal{D}_{zzx}$ correspond to the CDW order parameter with
$\left\langle \hat{C}_{\mathbf{k}, \sigma,\uparrow}^\dagger \hat{C}_{\mathbf{k}+2\mathbf{k}_c,\sigma,\uparrow} \right\rangle = \left\langle \hat{C}_{\mathbf{k}, \sigma,\downarrow}^\dagger \hat{C}_{\mathbf{k}+2\mathbf{k}_c,\sigma,\downarrow} \right\rangle$ and the SDW order parameter with
$\left\langle \hat{C}_{\mathbf{k}, \sigma,\uparrow}^\dagger \hat{C}_{\mathbf{k}+2\mathbf{k}_c,\sigma,\uparrow} \right\rangle = -\left\langle \hat{C}_{\mathbf{k}, \sigma,\downarrow}^\dagger \hat{C}_{\mathbf{k}+2\mathbf{k}_c,\sigma,\downarrow} \right\rangle$, respectively.

\begin{figure*}[!htbp]
   \centering
   \includegraphics[width=6.4in]{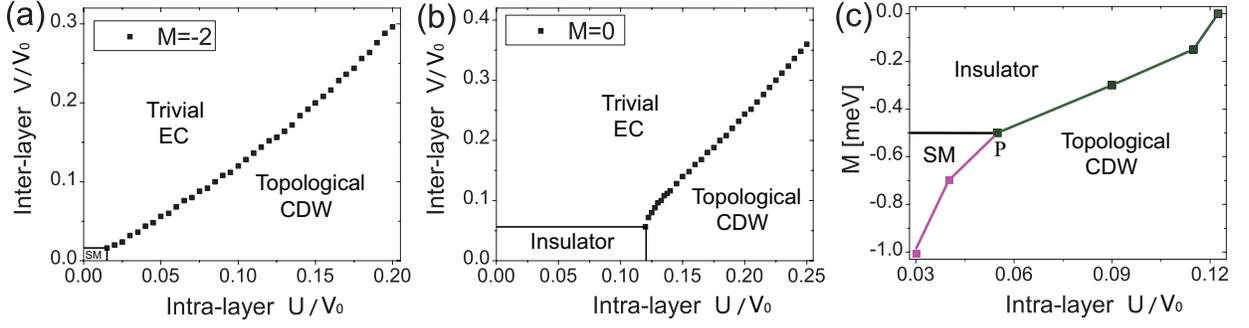}
   \caption{\label{fig-2-phase-diagram} The phase diagram is shown as a function of intra-layer interaction $U$ and inter-layer interaction $V$ for (a) $M=-2$ meV and (b) $M=0$ meV. (c) The phase diagram as function of $M$ and $U$ is shown for $V=0$.
   The pink line indicates the second order phase transition between semi-metal and Topological CDW, and the green line indicates the first order phase transition between insulator and Topological CDW. The parameters used are $B=-660$ meV$\cdot$nm$^{2}$, $A=37$ meV$\cdot$nm.}
\end{figure*}

{\it Topological CDW/SDW phase and trivial EC phase - }
The phase diagram of EC phase and CDW/SDW phases as a function of intra-layer interaction $U$ and
inter-layer interaction $V$ can be extracted from the self-consistent calculation of order parameters $\bar{\Delta}$ and free energy $\mathcal{F}(\bar{\Delta})$. We identify the parameter regimes for the order parameters $\mathcal{D}_{xxx}$ and $\mathcal{D}_{yxy}$, as well as $\mathcal{D}_{0zx}$ and $\mathcal{D}_{zzx}$, in the phase diagram with the numerical calculations \cite{budich_prl_2014}.
We only focus on EC order parameter $\mathcal{D}_{yxy}$ and CDW order parameter $\mathcal{D}_{0zx}$.
The phase diagrams are shown in Fig.~\ref{fig-2-phase-diagram} (a) and (b) for $M=-2$ meV and $M=0$,
respectively. For $M=-2$ meV, we start from semi-metal phase with electron and hole Fermi pockets in
the non-interacting limit. With increasing interactions, the EC phase occurs for a large inter-layer
interaction $V$ while the CDW phase emerges for a large intra-layer interaction $U$. The transition
from a semi-metal phase to the EC phase or CDW phase is of the second order nature while the transition
between the EC phase and the CDW phase is of the first order nature. Similar phase diagram is also
obtained for $M=0$. However, in this case, the system is in an insulating phase in the non-interacting
limit and as a result, a stronger interaction $U$ or $V$ is required to drive the system into
the CDW or EC phase. Numerically, we find the results for SDW order parameter is the same as CDW case and $\mathcal{D}_{xxx}$ is identical to $\mathcal{D}_{yxy}$. Another interesting issue is that for $M=0$, the transition from the insulating phase
to CDW phase is of the first order. As discussed in details in the appendix, there is a jump for the gap function $\bar{\Delta}$ around the transition point $U_c$. Furthermore, we find metastable states near $U_c$, with finite gap function $\bar{\Delta}$ but higher free energy $\mathcal{F}(\bar{\Delta})>\mathcal{F}(0)$. On the other hand, the transition from the semimetal phase to the CDW phase is of the second order. In Fig.~\ref{fig-2-phase-diagram} (c), we depict the phase diagram as a function of $M$ and $U$, in which a tri-critical point, labelled by ``P'',
is found on the phase diagram.

Next we explore the topological nature of EC phase and CDW phase by projecting
the Hamiltonian into the low energy subspace \cite{huba_book_2010} with the electron Fermi pocket around ${\bf k_c}$ and the hole Fermi pocket around ${\bf -k_c}$. Firstly, let us begin with CDW/SDW case, $\mathcal{H}_{\bar{\Delta}}=\left\lbrack\begin{array}{cc} \Delta_1 & 0 \\ 0 &\Delta_2 \end{array}\right\rbrack\otimes\tau_x$. By projecting out high energy bands shown as $\vert E,s-\mathbf{k}_c\rangle$ and $\vert H,s,\mathbf{k}_c\rangle$ in Fig. \ref{fig-1-sketch}(b), the Hamiltonian reads
\begin{align}\label{eq-eff-ham}
  \mathcal{H}_{\text{eff}}(\bar{\mathbf{k}}) = \left\lbrack\begin{array}{cc}
                      \epsilon_E s_0 - \mathcal{M}_1 & -\mathcal{D}_1 \\
                      -\mathcal{D}_1^\dagger & \epsilon_H s_0 - \mathcal{M}_2
                  \end{array}\right\rbrack ,
\end{align}
on the 4-by-4 low-energy basis $\left\{\vert E,s,\mathbf{k}_c\rangle, \vert H,s,-\mathbf{k}_c\rangle \right\}$, where the diagonal corrections are $ \mathcal{M}_1 =  \frac{\Delta_1 \Delta_1^\dagger}{\epsilon_E'} + \frac{A^2\vert\mathbf{k}_c\vert^2}{\epsilon_H'} $ and $ \mathcal{M}_2 =  \frac{\Delta_2^\dagger \Delta_2}{\epsilon_H'}  + \frac{A^2\vert\mathbf{k}_c\vert^2}{\epsilon_E'}$ with $\epsilon_E'=-4B\vert\mathbf{k}_c\vert^2$ and $\epsilon_H'=4B\vert\mathbf{k}_c\vert^2$. And the off-diagonal hybridization term is
\begin{align}\label{eq-eff-ham-d1-terms}
  \mathcal{D}_1 &= \frac{\Delta_1(Ak_x's_z - iAk_y's_0)}{\epsilon_E'} + \frac{(Ak_xs_z - iAk_ys_0)\Delta_2}{\epsilon_H'}
\end{align}
with $\mathbf{k}'=\bar{\mathbf{k}}-\mathbf{k}_c\sim-\mathbf{k}_c$ and $\mathbf{k}=\bar{\mathbf{k}}+\mathbf{k}_c\sim\mathbf{k}_c$. Firstly, the $\mathcal{D}_1$ term is given by $ A_{\text{eff}}(\bar{k}_xs_z - i\bar{k}_ys_0) $ for the topological CDW order parameter $\mathcal{D}_{0zx}$ ( $\Delta_1=-\Delta_2 = \Delta_0 s_0$ ), while for the topological SDW order parameter $\mathcal{D}_{zzx}$ ($\Delta_1=-\Delta_2 = \Delta_0 s_z$),
this term becomes $ A_{\text{eff}}(\bar{k}_xs_0 - i\bar{k}_ys_z) $.
We notice that the above Hamiltonian reproduces the standard form of BHZ model with the renormalized
mass term $\tilde{M} \sim M - \frac{\Delta_0^2-A^2\vert\mathbf{k}_c\vert^2}{\epsilon_E'}$
and linear term coefficient $A_{\text{eff}}\approx \frac{2A\Delta_0}{\epsilon_E'}$.
As a consequence, we expect that the system is in the quantum spin Hall state with helical edge
transport. It is interesting to notice that when the condition $\Delta_0^2>A^2\vert\mathbf{k}_c\vert^2$
is satisfied, the renormalized mass term $\tilde{M}$ becomes inverted even we start from a normal
mass $M>0$.
As for the trivial EC case, we only need transform Eq.~\eqref{eq-ham-mf} to $\mathcal{H}_{\text{eff}}(\bar{\mathbf{k}}) = \left\lbrack \begin{array}{cc} \epsilon_E s_0 & \mathcal{D}_3 \\ \mathcal{D}_3^\dagger & \epsilon_H s_0\end{array} \right\rbrack$ with $\mathcal{D}_3= \Delta_0 s_x$ or $\Delta_0  s_y$ in the first order perturbation level, giving rise to a full trivial gap by the EC phase.

\begin{figure}[!htbp]
   \centering
   \includegraphics[width=3.3in]{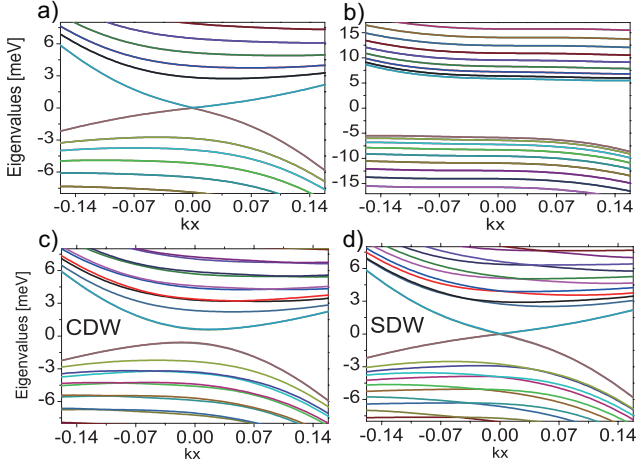}
   \caption{\label{fig-3-add-helical-edge-states} Illustration of helical edge states in the reduced and folded Brillouin zone without Zeeman term in a) CDW/SDW order parameter and b) EC order parameter, and with Zeeman term ($g_e=g_h=1$ meV) in c) CDW order parameter and d) SDW order parameter. Other parameters are $M=0, A\phi_0=9.4, B\phi_0^2=-2.5$ with $\phi_0=\pi/20$ and $\Delta_0=12$ in unit of meV.}
\end{figure}

To confirm the above conclusion that CDW/SDW phase is topologically non-trivial while EC phase is topologically
trivial, we perform a direct calculation of energy dispersion in a slab configuration with an open boundary condition \cite{shen_ti_book_2013} to reveal helical edge states. Indeed, as shown in both Fig.~\ref{fig-3-add-helical-edge-states}(a)(b), we find two counter-propagating modes in the CDW/SDW phase and a full insulating gap in the EC phase. For topological CDW phase, we estimate the effective hybridization term $A_{\text{eff}} \sim 144$ meV and the renormalized inversion gap $\tilde{M} \sim -5.4$ meV for the low-energy effective theory, thus consistent with the edge state calculation.

In the above discussion, the Zeeman effect has not been taken into account, which can hybridize the
opposite spin block in Eq.~\eqref{eq-ham-mf} and thus may destroy the helical edge transport.
The in-plane Zeeman effect takes the form
$\mathcal{H}_{Z,e/h} = g_{e/h} \mu_B B_y s_y$ with the Bhor magneton $\mu_B$, the magnetic field $B_y$ along the y direction and the g-factor $g_e$ and $g_h$ for electron $\vert E,s\rangle$ band and hole $\vert H,s\rangle$ band
\cite{mu_apl_2016,karalic_arxiv_2016}, respectively.
We project the Zeeman term into the basis of low energy bands
$\left\{\vert E,s,\mathbf{k}_c\rangle, \vert H,s,-\mathbf{k}_c\rangle \right\}$
and find that the corrections to the Zeeman term are given by
\begin{align}\label{eq-g-factor-correction}
	\Delta\mathcal{H}_{Z,e} &=  \frac{\mu_B g_hB_y\vert A\mathbf{k}_c\vert^2}{(\epsilon_H')^2-(\mu_Bg_hB_y)^2} s_y
	\nonumber\\
	&+ \frac{\mu_Bg_eB_y}{(\epsilon_E')^2-(\mu_B g_eB_y)^2}\Delta_1 s_y \Delta_1^\dagger , \nonumber \\
	\Delta\mathcal{H}_{Z,h} &=  \frac{\mu_Bg_eB_y\vert A\mathbf{k}_c\vert^2}{(\epsilon_E')^2-(\mu_B g_eB_y)^2} s_y
	\nonumber \\
	&+ \frac{\mu_B g_hB_y}{(\epsilon_H')^2-(\mu_B g_hB_y)^2}\Delta_2^\dagger s_y \Delta_2
\end{align}
for electron bands and hole bands, respectively.
There are two terms in the corrections of the Zeeman coupling.
The first term is due to the low energy physics that occurs at the finite
momentum $\pm {\bf k_c}$, at which the electron and hole bands are hybridized with each other,
while the second term directly comes from the influence of CDW/SDW order parameters. This Hamiltonian
clearly shows that the g-factor of the Zeeman term is strongly renormalized by interactions.
Intriguingly, from the numerical calculation based on realistic parameters for the four band model \cite{franz_book_2013},
we find helical edge modes are robust in the SDW phase but destroyed in the CDW phase
when $g_e\approx g_h$,
as shown in Fig.~\ref{fig-3-add-helical-edge-states}(c) and (d).
These features of helical edge modes under the influence of Zeeman effect can be qualitatively
understood from the perturbation results (Eq.\ref{eq-g-factor-correction}),
as discussed in the appendix.
These results reveal the importance of interaction correction for Zeeman effect, but
we emphasize that they are material dependent.

{\it Discussion and conclusion -}
In this work, we have shown that the interaction effect can drive the InAs/GaSb quantum wells from
a semi-metal phase or an trivial insulating phase into a topologically non-trivial CDW/SDW phase
or a trivial EC phase under an in-plane magnetic field.
Our results suggest that topological CDW/SDW phase might be the underlying physical reason
for the robust quantum spin Hall state that was observed in InAs/GaSb quantum wells under
an in-plane magnetic field \cite{du_arxiv_2015}. Furthermore, we find that the g-factor of the Zeeman effect
is also significantly renormalized under interaction and may be reduced in certain parameter regime.
Experimentally, the out-of-plane g-factor was known to be around $10$, reported in Ref.~\cite{mu_apl_2016,karalic_arxiv_2016} for electron bands of InAs/GaSb quantum well systems.
CDW/SDW phase might be experimentally probed through the pinning-depinning
transition \cite{rmp_gruner_1998} or some interference phenomena \cite{book_gruner_2000}.
We also notice recent debates about the nature of edge modes in InAs/GaSb quantum wells \cite{nichele_njp_2016,nguyen_prl_2016}
and our proposal might provide additional information for this issue.
Topological CDW/SDW phase is different from the previous discussed topological EC phase since
it only emerges at a strong in-plane magnetic field, which is the valid regime of our discussion here.
Nevertheless, it may also be related to the topological EC phase \cite{pikulin_prl_2014} with p-wave type EC order parameter, because the second-order off-diagonal term in Eq.~\eqref{eq-eff-ham-d1-terms} induced by CDW/SDW order parameter may also be regarded as p-wave type. Since our topological CDW/SDW phase is only valid in a strong magnetic field while p-wave topological EC phase can exist at zero magnetic field, it is an interesting question to ask how these two phases are connected in a small magnetic field regime, which deserves a future study and is beyond the scope of the current paper.

LHH and CXL would like to thank Jan Carl Budich and Paolo Michetti for useful discussions in related physics and numerical calculations. LHH would also like to thank Wei-Qiang Chen, Zewei Chen and Dong-Hui Xu for helpful discussions.
C.-X.L. acknowledge the support from Office of Naval Research
(Grant No. N00014-15-1-2675).
FCZ is supported by National Basic Research Program of China (No.2014CB921203) and NSFC (No.11674278).
YZ is supported by National Key Research and Development Program of China (No.2016YFA0300202), National Basic Research Program of China (No.2014CB921201), NSFC (No.11374256) and the Fundamental Research Funds for the Central Universities in China.

\bibliographystyle{apsrev4-1}
\bibliography{Reference_main}

\end{document}